\documentclass[twocolumn,showpacs,floatfix,superscriptaddress,amsmath,amssymb,prl]{revtex4}
\usepackage{amsmath}
\usepackage{}
\usepackage{mathrsfs}
\usepackage{txfonts}
\usepackage{amssymb}
\usepackage{graphicx}
\usepackage{hyperref}
\usepackage{ulem}
 \usepackage{overpic}
 \usepackage{psfrag}
\usepackage{tabularx}
\usepackage{array}
\usepackage{placeins}
\usepackage{color}
\usepackage{subfigure}

\makeatletter

\begin{document}
\title{Symmetry-protected trivial phases and quantum phase transitions in an anisotropic antiferromagnetic spin-1 biquadratic model}

\author{Xi-Hao Chen}
\affiliation{Centre for Modern Physics, Chongqing University,
Chongqing 400044, The People's Republic of China}

\affiliation{Research Institute for New Materials and Technology, Chongqing University of Arts and Sciences, Chongqing 400000, The People's Republic of China}

\author{Ian McCulloch}
\affiliation{Department of Physics, The University of Queensland, Brisbane, Australia}

\author{Murray T. Batchelor}
\affiliation{Department of Theoretical Physics, Research School of Physics, The Australian National University, Canberra ACT 2601, Australia}
\affiliation{Mathematical Sciences Institute, The Australian National University, Canberra ACT 2601, Australia}
\affiliation{Centre for Modern Physics, Chongqing University,
Chongqing 400044, The People's Republic of China}

\author{Huan-Qiang Zhou}
\affiliation{Centre for Modern Physics, Chongqing University,
Chongqing 400044, The People's Republic of China}

\begin{abstract}
The ground state phase diagram is obtained for an antiferromagnetic  spin-1 anisotropic biquadratic model. 
With the help of symmetry and duality transformations, three symmetry-protected trivial phases and one dimerized symmetry breaking phase are found. 
Local and nonlocal order parameters are identified to characterize these phases.
Quantum phase transitions between the symmetry-protected trivial phases belong to the Gaussian universality class with central charge $c=1$, 
and quantum phase transitions from the symmetry-protected trivial phases to the dimerized phase belong to the Ising universality class with central charge $c=1/2$. 
In addition, the model admits three characteristic lines of factorized ground states, 
which are located in the symmetry-protected trivial phases instead of a symmetry breaking phase, in sharp contrast to other known cases.
\end{abstract}

\pacs{75.10.Jm, 64.70.Tg, 71.10.Hf, 75.10.Pq}

\maketitle

{\it Introduction.}--Much attention has been focused on critical phenomena in quantum many-body systems, 
with an aim towards a complete classification of quantum states of matter.  
In this regard, significant progress has been made for quantum spin systems in one dimension, 
resulting in the introduction of novel concepts, such as symmetry-protected topological order~\cite{wenxg}, 
and symmetry-protected trivial (SPt) order~\cite{Pollmann2}.
A SPt phase  is a
symmetric phase connected adiabatically to a product state, and is characterized in terms of a non-local order parameter defined by
the combined operation of the
site-centered inversion symmetry with a $\pi$-rotation around the $y$-axis in the spin space.
As a consequence, such a SPt phase is different from a symmetry-protected topological phase~\cite{wenxg}.  
Therefore, it is expected to also play an important role in classifying quantum states of matter~\cite{Pollmann2,orus}.
However, it remains unclear whether or not the current characterization of SPt phases in terms of the non-local order parameter is generic enough for any possible SPt phases.

On the other hand, dualities, symmetries and factorized ground states combine to  
play a significant role in characterizing physical properties for quantum many-body systems~\cite{Zhou}.  
An intriguing question is how then to demonstrate the powerfulness of all these concepts in one single illustrative example.
In this work we investigate the nature of SPt phases appearing in the  ground state phase diagram for an anisotropic generalization 
of the spin-1 biquadratic model ~\cite{Murray,Klumper}, which is known to be in a dimerized phase. 
The more general model proposed here is described by the Hamiltonian
\begin{eqnarray}\label {ham}
H=-\sum_{i=-\infty}^{\infty} \left(J_{x} \, S^{x}_{i}{S^{x}_{i+1}}+J_{y} \, S^{y}_{i}{S^{y}_{i+1}}+J_{z} \, S^{z}_{i}S^{z}_{i+1}\right)^{2},
\end{eqnarray}
where $S_{i}^{\alpha}$ ($\alpha$ = $x$, $y$ and $z$) denote spin-1 operators at site $i$ on an infinite-size chain, with 
$J_{x}$, $J_{y}$ and $J_{z}$ the spin couplings describing the interactions between the $x$, $y$ and $z$ components.
The model is antiferromagnetic, in the sense that it becomes the antiferromagnetic spin-$1/2$ XYZ model, 
which itself is an exactly solvable model~\cite{Baxter}, if $S_{i}^{\alpha}$ are spin-$1/2$ operators.

The model Hamiltonian (\ref{ham}) reduces to the spin-1 SU(2)-invariant biquadratic model when $J_x=J_y=J_z$. 
The spin-1 biquadratic model has been extensively investigated. 
It can be either mapped to the 9-state Potts model~\cite{Murray} or solved directly in terms of Bethe Ansatz or functional relations~\cite{Klumper}.
It is thus known to be in a dimerized phase with a relatively small spectral gap. 
The Hamiltonian (\ref{ham}) of the anisotropic generalization of this model  commutes with the three operators 
$\Sigma_x = \Sigma_{i}(-1)^i ((s_i^{y})^2-(s_i^z)^2)$, 
$\Sigma_y = \Sigma_{i}(-1)^i ((s_i^{z})^2-(s_i^x)^2)$ and $\Sigma_z = \Sigma_{i}(-1)^i ((s_i^{x})^2-(s_i^y)^2)$. 
In particular, these three commuting operators, combining with $S_i^{\alpha}$, generate an SU(3) symmetry when $J_x = J_y = J_z$~\cite{affleck}.  
It is the staggered nature of the symmetry operators which explains why spontaneous dimerization occurs in the spin-1 SU(2)-invariant biquadratic model~\cite{Murray}.

{\it Duality transformations.}--Key information about the nature of the phase diagram can be obtained from duality relations among the spin couplings. 
Here quantum duality is represented by a local unitary transformation $U$ acting 
on Hamiltonian (\ref{ham}).
For convenience and simplicity we define the variables $X=J_x/J_z$ and $Y= J_y/J_z$ and consider $H(X,Y)$.
In general if $H(X^\prime,Y^\prime)$ is dual to $H(X,Y)$ there should exist a unitary transformation $U$ satisfying 
\begin{equation}
H(X,Y)=k(X,Y) \,U \, H(X^\prime,Y^\prime) \, U^{\dag}.
\label{gen}
\end{equation}
The coupling parameters $X^\prime$ and $Y^\prime$ are 
functions of $X$ and $Y$ with $k(X,Y)$ being positive.

For Hamiltonian (\ref{ham}) 
there are two symmetry transformations and four duality transformations, as presented in Table I.
These symmetry and duality transformations imply six self-dual lines defined by 
$X= \pm 1$, $Y= \pm 1$ and $Y/X= \pm 1$.

\begin{table}[t]
\caption{\label{tab:example} The six symmetry and duality transformations for the biquadratic Hamiltonian (\ref{ham}) defined in relation (\ref{gen}).}
\begin{ruledtabular}
\begin{tabular}{lccrrr}
  $U$ & spin transformation & $k(X,Y)$ & $X^\prime$ & $Y^\prime$ \\
\hline
  $U_a$ & $S_{i}^{x} \rightarrow -S_{i}^{y}$, $S_{i}^{y} \rightarrow -S_{i}^{x}$, $S_{i}^{z} \rightarrow -S_{i}^{z}$,   & 1 & $-Y$ & $-X$ \\ 
& $S_{i+1}^{x} \rightarrow S_{i+1}^{y}$, $S_{i+1}^{y} \rightarrow S_{i+1}^{x}$, $S_{i+1}^{z} \rightarrow -S_{i+1}^{z}$ & & & \\
  $U_b$ & $S_{i}^{x} \leftrightarrow S_{i}^{y}$, $S_{i}^{z} \rightarrow -S_{i}^{z}$  & 1 & $Y$ & $X$ \\
  $U_1$ & $S_{i}^{x} \rightarrow -S_{i}^{x}$, $S_{i}^{y} \rightarrow S_{i}^{z}$, $S_{i}^{z} \rightarrow S_{i}^{y}$  & $Y^2$ & $X$ & $1/Y$ \\
 $U_2$ & $S_{i}^{x} \rightarrow -S_{i}^{x}$, $S_{i}^{y} \rightarrow -S_{i}^{z}$, $S_{i}^{z} \rightarrow -S_{i}^{y}$, & $Y^2$ & $-X/Y$ & $-1/Y$ \\
 & $S_{i+1}^{x} \rightarrow -S_{i+1}^{x}$, $S_{i+1}^{y} \rightarrow S_{i+1}^{z}$, $S_{i+1}^{z} \rightarrow S_{i+1}^{y}$ & & & \\
  $U_3$ & $S_{i}^{x} \rightarrow S_{i}^{z}$, $S_{i}^{y} \rightarrow -S_{i}^{y}$, $S_{i}^{z} \rightarrow S_{i}^{x}$ & $X^2$ & $1/X$ & $Y/X$ \\
  $U_4$ & $S_{i}^{x} \rightarrow -S_{i}^{z}$, $S_{i}^{y} \rightarrow -S_{i}^{y}$, $S_{i}^{z} \rightarrow -S_{i}^{x}$, & $X^2$ & $-1/X$ & $-Y/X$ \\
 & $S_{i+1}^{x} \rightarrow S_{i+1}^{z}$, $S_{i+1}^{y} \rightarrow -S_{i+1}^{y}$, $S_{i+1}^{z} \rightarrow S_{i+1}^{x}$ & & & \\
 \end{tabular}
\end{ruledtabular}
\end{table}

\begin{figure}[t]
\includegraphics[width=0.30\textwidth]{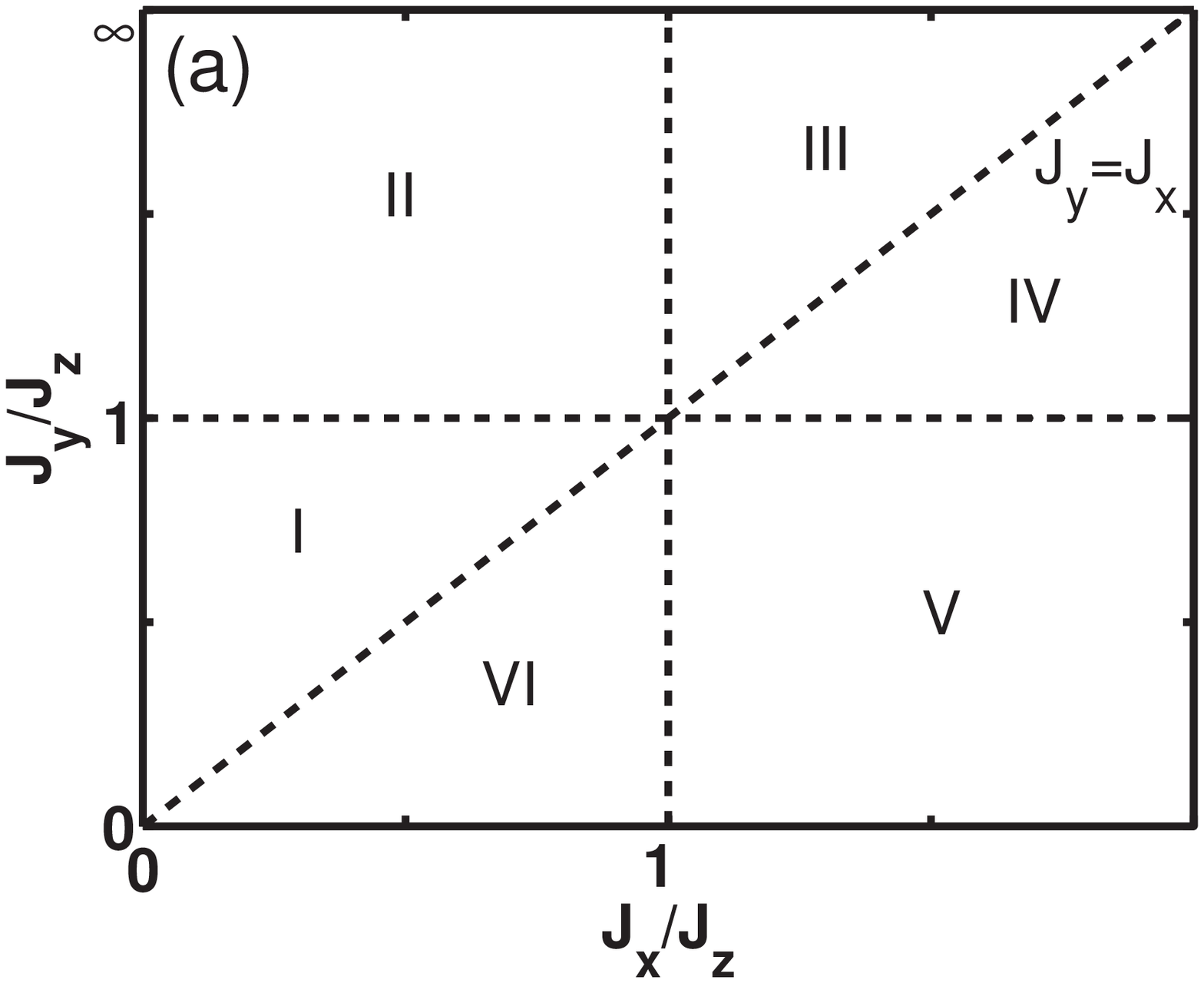}
\includegraphics[width=0.30\textwidth]{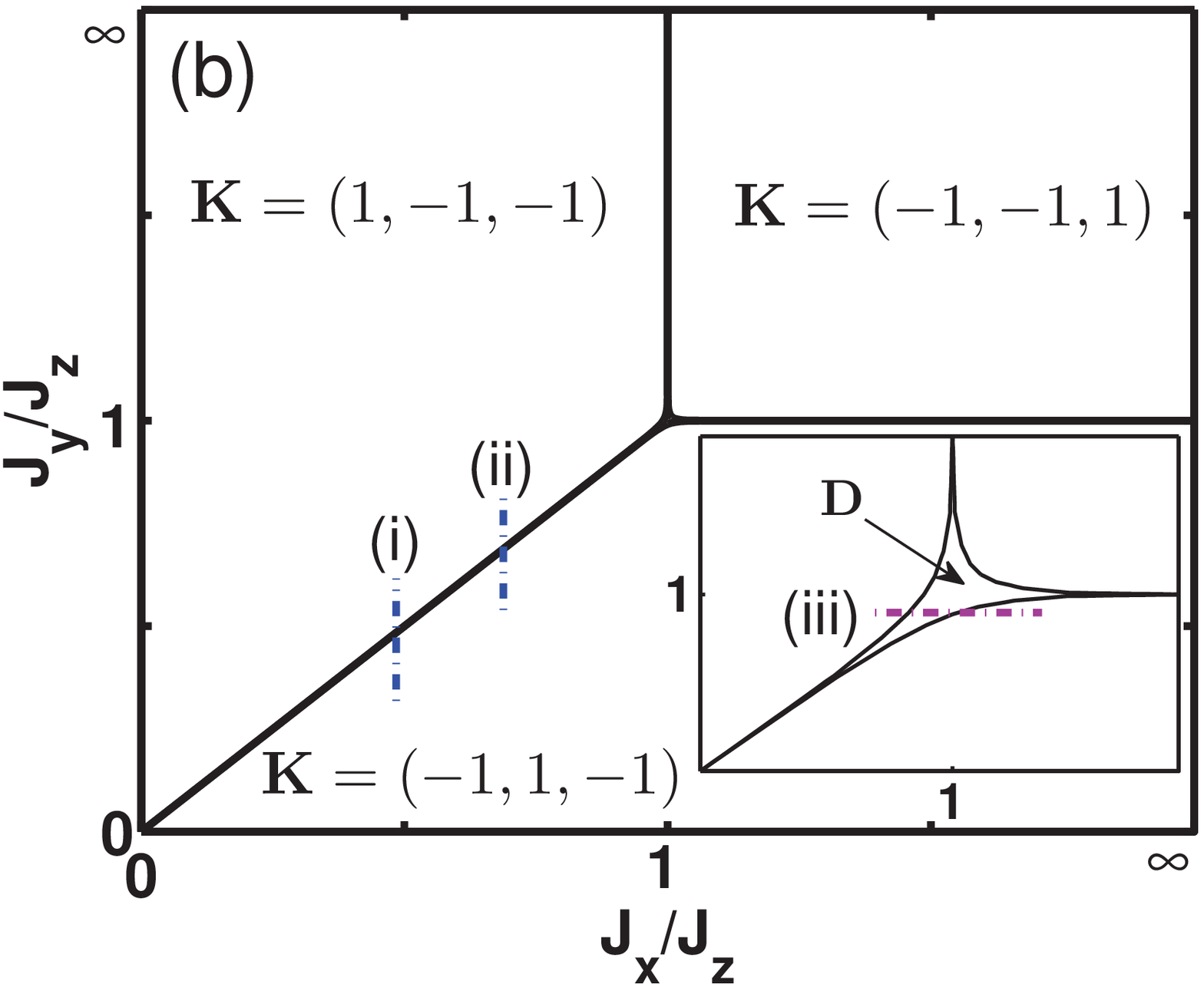}
\caption{
(a) The six dual regions in the $J_{x}/J_{z} \ge 0$ and $J_{y}/J_{z} \ge 0$ parameter space of the 
spin-1 biquadratic anisotropic model. Region I is the principal regime.
(b) The phase diagram characterized by three $Z_{2}$ combined symmetry operations 
${\boldsymbol K}=(K_{x},K_{y},K_{z})$ and order parameters ${\boldsymbol D}=(D_{x},D_{y},D_{z})$.
The inset shows a magnification of the tiny region defining the dimerized phase.
The dashed-dotted paths labeled by (i), (ii) and (iii) are sample lines discussed in the text to 
show how to characterize these phases.
Both the horizontal and vertical axes are shown to tan inverse scale.
}
\label{fig1duality}
\end{figure}

{\it Ground state phase diagram.}--To obtain the phase diagram we consider the 
self-dual lines $J_{x}/J_{z}= 1$, $J_{y}/J_{z}= 1$ and $J_{y}/J_{x}= 1$, 
which delineate the six regions shown in Fig.~\ref{fig1duality}(a).
Because of the various symmetries and dualities, we only need to focus on one of these six regions, with 
the whole phase diagram following by mapping with the help of the duality transformations.
We thus define region I as the principal regime.
To determine the phase boundary in region I,
we examine order parameters based on numerical simulations in terms of the infinite Time-Evolving Block Decimation (iTEBD)  algorithm~\cite{vidal} and the infinite Density Matrix Renormalization Group (iDMRG) algoritnm~\cite{Ian}.  In both algorithms, ground state wave functions are represented in terms of infinite Matrix Product States (iMPS). 
We will also make use of the von Neumann entropy.
Fig.~\ref{fig1duality}(b) shows the ground state phase diagram determined in this way, as discussed in detail below.

{\it Factorized states.}--When $J_{x}=0$ the system is in a factorized ground state.
The wave function of this factorized state can be written as a simple product of the vector $\frac{1}{\sqrt{2}}(1,0,-1)$.
The corresponding factorized energies are given by $e=-(J_{y}^{2}+ J_{z}^{2})$.
Similarly, the system is also in a factorized state when $J_{y}=0$ or $J_z=0$, respectively, as follows from the duality transformations. 
The wave functions can then be written as products of the vector $\frac{1}{\sqrt{2}}(1,0,1)$ or $(0,1,0)$
with energies $e=-(J_{x}^{2}+J_{z}^{2})$ or $e=-(J_{x}^{2}+J_{y}^{2})$, respectively.
These three factorized states are dual to each other.
 Note that, if we choose $J_z$ as an energy scale, then $J_z =0$ is equivalent to saying that both $J_x$ and $J_y $
 approach infinity with a fixed ratio $J_x/J_y$.
We also note that when $J_{x}=J_{y}=0$, the model has $2^{N}$ degeneracies with entanglement
varying between $0$ and $\infty$, consistent with previous results for systems with largely degenerate ground states \cite{Doyon}.

{\it Symmetry-protected trivial phases.}--Spontaneous symmetry breaking (SSB) occurring in quantum many-body systems 
implies the existence of local order which can be characterized by local order parameters. 
Significantly, there exist other concepts of order in quantum many-body systems which are beyond Landau theory due to the absence of any SSB.
Examples are the three SPt phases mentioned above, which cannot be successfully characterized by local order parameters.
Fortunately appropriate non-local order parameters can be used to characterize such order, which may be regarded as a natural extension of  the SPt order introduced in Ref.~\cite{Pollmann2}. 
Indeed, distinct non-local order parameters can successfully distinguish various symmetry-protected phases
and are also effective for symmetry-protected gapped phases with partial
symmetry breaking \cite{Haegeman,Schuch}.

For a given ground state wave function $|\psi\rangle$ of an infinite-size spin chain represented by 
the iMPS, the site-centered non-local order parameters can be written based on
reversing an odd-sized segment of the chain and then calculating the resulting overlap \cite{Pollmann2,Pollmann}.  
This can be defined in terms of the inversion operator 
\begin{eqnarray}\label{s2lplus1}
O_{L}^{\alpha}=\frac{\langle \psi\mid {I_{(1,L)}}\Sigma^{\alpha}_{(1,L)} \mid \psi \rangle}{tr(\lambda_{A}^{2}\lambda_{B}^{2})},
\end{eqnarray}
where $O_{L}^{\alpha}$ is the inversion on the segment from $1$ to $L$
with internal symmetry operations $\Sigma^{\alpha}_{(1,L)}$ acting on the physical indices.
Here $\Sigma^{\alpha}_{(1,L)}$ is $\exp(i\pi S^{\alpha})$, with $S^{\alpha}$ a spin-1 matrix and $\alpha=x, y$ or $z$.
The segment length $L$ is odd.
$\lambda_{A}$ and $\lambda_{B}$ are Schmidt decomposition coefficients.

\begin{figure}[t]
\includegraphics[angle=270,width=0.45\textwidth]{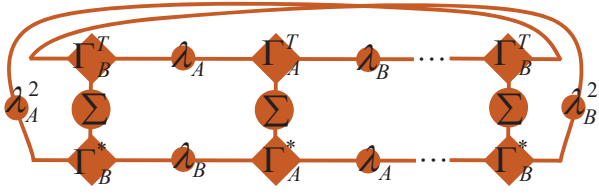}
\caption{
Graphical representation of non-local order parameters in the MPS picture. 
$\Gamma_{A}$ and $\Gamma_{B}$ are three-index tensors; $\lambda_{A}$ and $\lambda_{B}$ are Schmidt decomposition coefficients.
These four tensors with a two-site translation invariance form the ground state wave function $|\psi\rangle$ of the infinite-size spin chain.
}
\label{fig2Nonlocalorderparameter}
\end{figure}

This definition respects two-site translation invariance due to the fact that
it can work on non-SSB and SSB wave functions directly.
In the absence of SSB, $O_{L}^{\alpha}$ gives a $\pm 1$ value.
Conversely, $O_{L}^{\alpha}$ gives a $0$ value for SSB.
Fig.~\ref{fig2Nonlocalorderparameter} shows the graphical representation
of non-local order parameters in the MPS framework.
Each SPt phase can then be characterized by the values $(O_{L}^{x},O_{L}^{y},O_{L}^{z})$ of the non-local order parameters.  
When $L$ approaches the thermodynamic limit, the non-local order parameters $(O_{L}^{x},O_{L}^{y},O_{L}^{z})$
obtain the exact values $(K_{x},K_{y},K_{z})$. 
We denote these three $Z_{2}$ combined symmetry operations by ${\boldsymbol K} = (K_{x},K_{y},K_{z})$.

\begin{figure}[t]
 \centering
\includegraphics[width=0.35\textwidth]{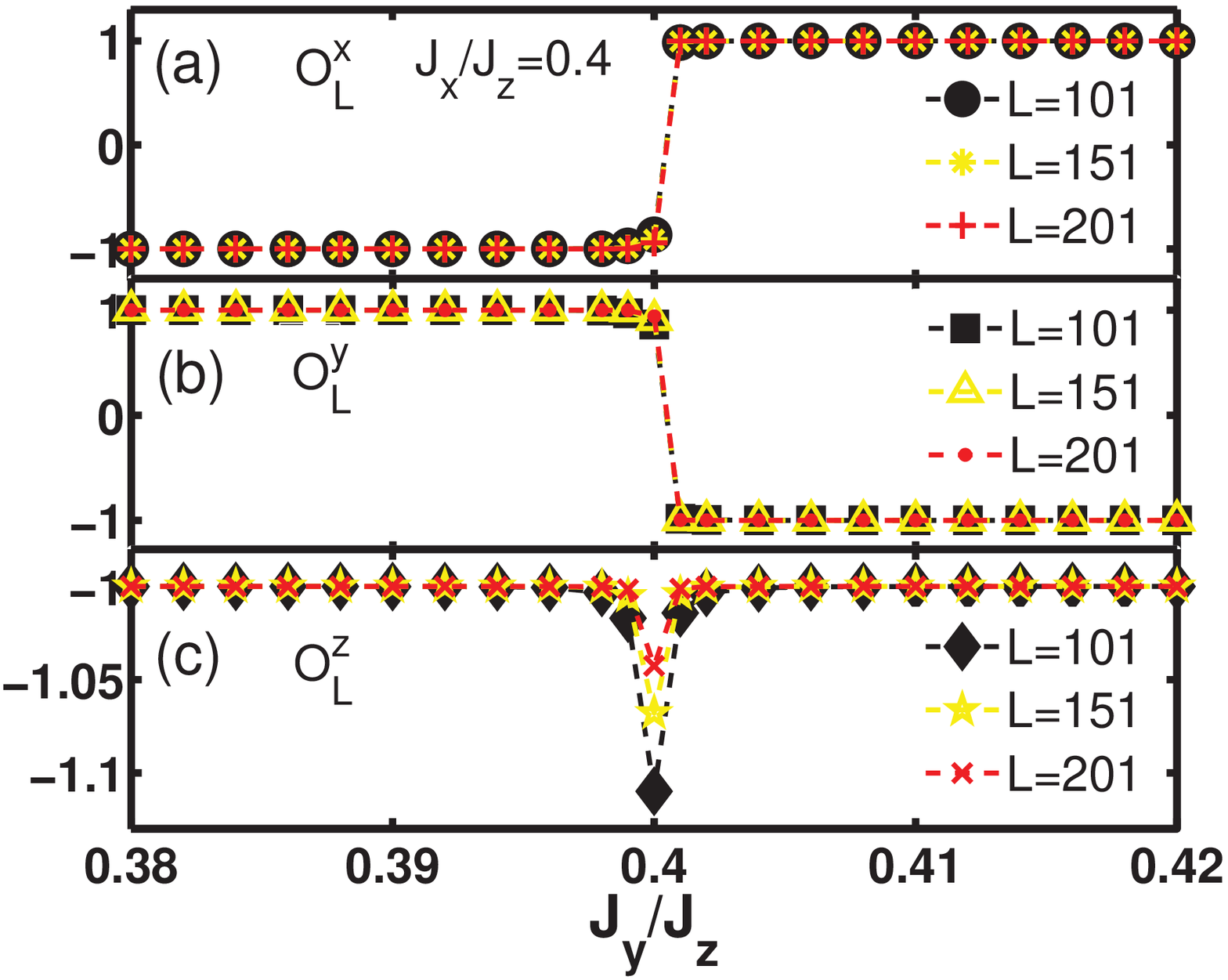}
\includegraphics[width=0.35\textwidth]{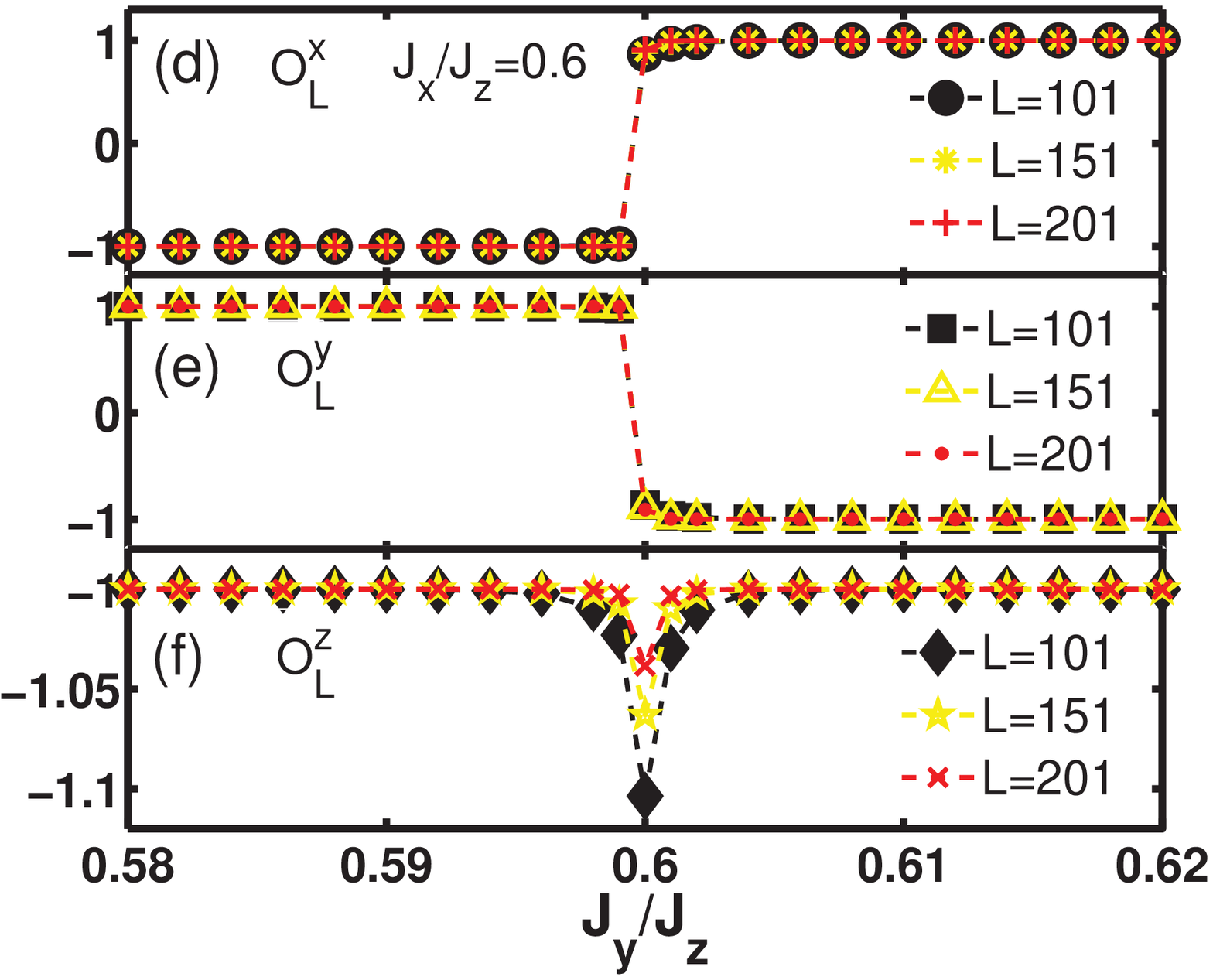}
\caption{
Non-local order parameters $O_{L}^{\alpha}$ as
a function of coupling $J_{y}/J_{z}$ with
truncation dimension $\chi=100$ for fixed $J_{x}/J_{z}=0.4$ (top) and $0.6$ (bottom).
The parameter $L$ shown is the inversion block size.
QPT points are identified as $J_{y}^{c}/J_{z}=0.4$ and $0.6$.
For each panel the symmetry parameter values ${\boldsymbol K}=(-1,1,-1)$ for the left side and ${\boldsymbol K}=(1,-1,-1)$ for the right side of the QPT point.
The data sets correspond to the lines (i) and (ii) in Fig.~\ref{fig1duality}(b).
}
\label{fig3Nonlocal04}
\end{figure}

We employ the definition of non-local order parameters (\ref{s2lplus1}) on the three sample lines, (i) $J_{x}/J_{z}=0.4$,
(ii) $J_{x}/J_{z}=0.6$ and (iii) $J_{y}/J_{z}=0.988$, as indicated in the phase diagram Fig.~\ref{fig1duality}(b).
First consider lines (i) and (ii).
In Fig.~\ref{fig3Nonlocal04}, we plot
$O_{L}^{x}$, $O_{L}^{y}$ and $O_{L}^{z}$ as functions of $J_{y}/J_{z}$
for fixed $J_{x}/J_{z}=0.4$ and fixed $J_{x}/J_{z}=0.6$ with truncation dimension $\chi=100$.
Quantum phase transition (QPT) points are located at $J_{x}^{c}/J_{z}=0.4$ and $J_{x}^{c}/J_{z}=0.6$, respectively.
Increasing the inversion block size $L$ from $L=101$ to $L=201$,
$O_{L}^{x}$ reaches the saturation value $-1$ most efficiently when $J_{y}/J_{z}$ is away from the QPT point $J_{y}^{c}/J_{z}$.
In contrast, $O_{L}^{x}$ saturates much slower in the vicinity of $J_{y}^{c}/J_{z}$.
The values of $O_{L}^{y}$ and $O_{L}^{z}$ behave similarly.

In this way the SPt phase on the right hand side of QPT points $J_{y}^{c}/J_{z}$
is characterized by ${\boldsymbol K}=(1,-1,-1)$, 
which corresponds to the SPt phase above (PA) diagonal phase boundaries.
Similarly, the phase on the left hand side of QPT points $J_{y}^{c}/J_{z}$ is  
characterized by ${\boldsymbol K}=(-1,1,-1)$,
which corresponds to the SPt phases below (PB) diagonal phase boundaries.
Based on the dual regions in Fig.~\ref{fig1duality}(a), the 
SPt PA phase is in region I and the SPt PB phase is in region VI.
These two phases are thus dual to each other, separated by the self-dual line $J_{y}= J_{x}$.
The values ${\boldsymbol K}=(1,-1,-1)$ of the non-local order parameters 
in region I and ${\boldsymbol K}=(-1,1,-1)$ in region VI foretell their duality.
Based on duality, one knows, e.g., that the non-local order parameters of the SPt phase in region III and
IV can be written as ${\boldsymbol K}=(-1,-1,1)$.

\begin{figure}[b]
\includegraphics[width=0.35\textwidth]{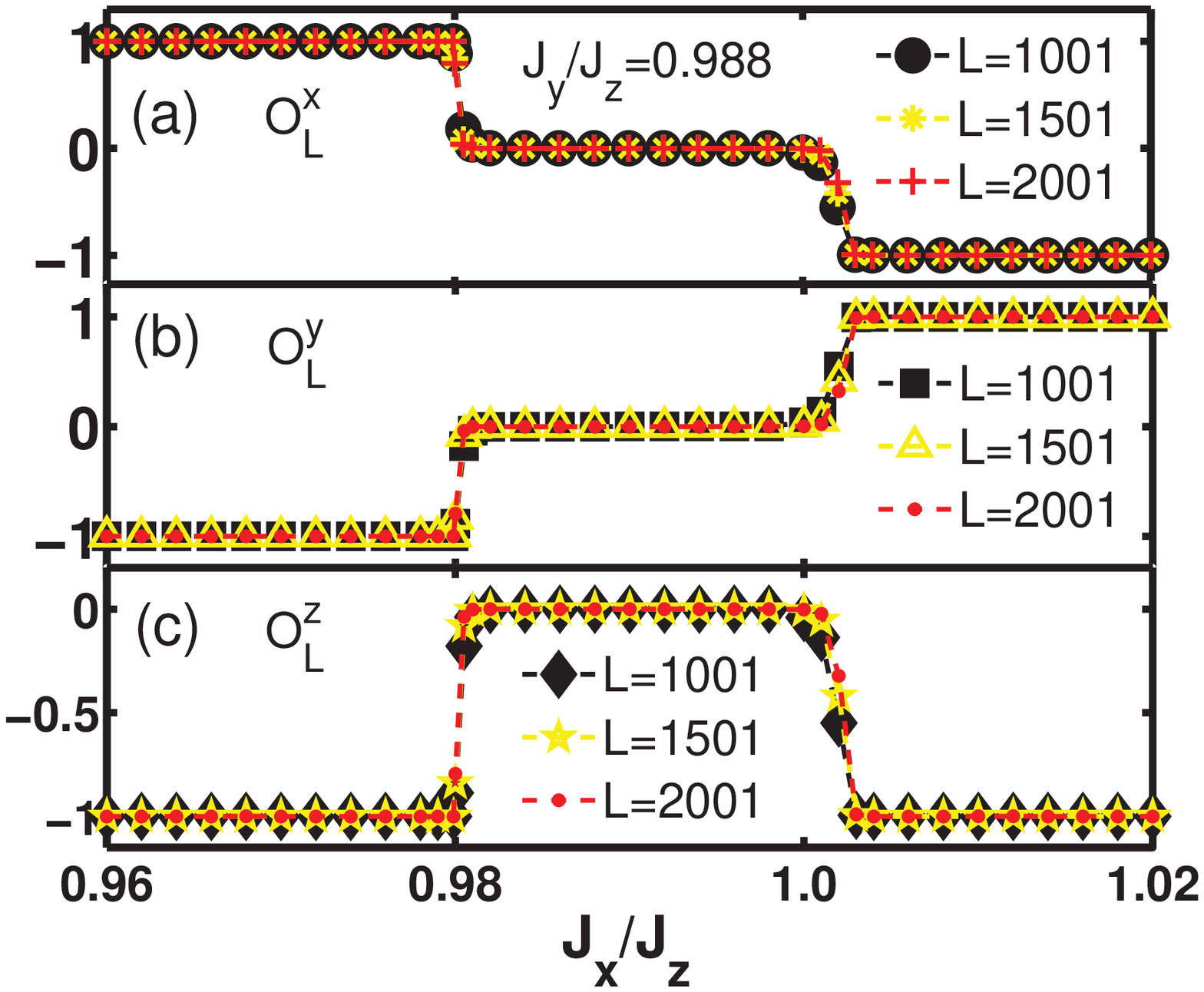}
\vskip 1mm
\includegraphics[width=0.35\textwidth]{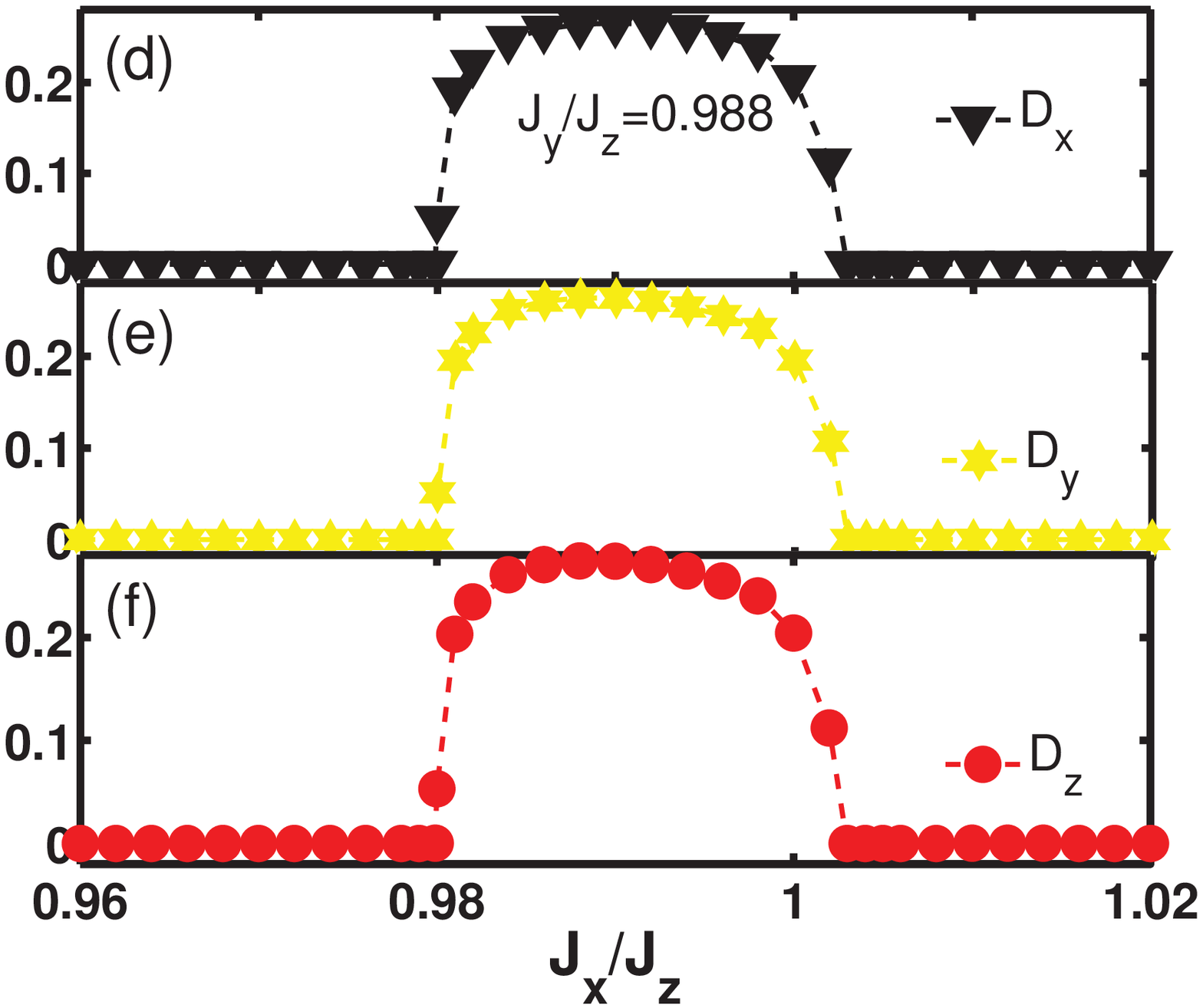}
\caption{
(top) Non-local and (bottom) local order parameters as
a function of control parameter $J_{x}/J_{z}$ with
truncation dimension $\chi=100$ for fixed $J_{y}/J_{z}=0.988$.
Two QPT points are detected at $J_{x}^{c_{1}}/J_{z}=0.9799$ and $J_{x}^{c_{2}}/J_{z}=1.003$.
Here ${\boldsymbol K}=(1,-1,-1)$ on the left hand side of $J_{x}^{c_{1}}/J_{z}$ and ${\boldsymbol K}=(-1,1,-1)$ on the right hand side of $J_{x}^{c_{2}}/J_{z}$. 
In the region between $J_{x}^{c_{1}}/J_{z}$ and $J_{x}^{c_{2}}/J_{z}$, ${\boldsymbol K}=(0,0,0)$,
where the dimerized phase is characterized by the combined local order parameter ${\boldsymbol D} =(D_{x},D_{y},D_{z})$.
The data sets correspond to the line (iii) in Fig.~\ref{fig1duality}(b).
}
\label{fig4Nonlocal0988}
\end{figure}

Turning to line (iii) in Fig.~\ref{fig1duality}(b) we plot $O_{L}^{x}$, $O_{L}^{y}$ and
$O_{L}^{z}$ as a function of $J_{x}/J_{z}$
for fixed $J_{y}/J_{z}=0.988$ with truncation dimension $\chi=100$ in Fig.~\ref{fig4Nonlocal0988}.
Here two QPT points are located at $J_{x}^{c_{1}}/J_{z}=0.9799$ and $J_{x}^{c_{2}}/J_{z}=1.003$.
The SPt phase on the left hand side of $J_{x}^{c_{1}}/J_{z}$ corresponding to PA
is characterized by ${\boldsymbol K}=(1,-1,-1)$.
The SPt phase on the right hand side of $J_{x}^{c_{2}}/J_{z}$ corresponding to PB
is characterized by ${\boldsymbol K}=(-1,1,-1)$.
We note that the saturation rate of $O_{L}^{x}$, $O_{L}^{y}$ and $O_{L}^{z}$
at $J_{x}^{c_{1}}/J_{z}=0.9799$ and $J_{x}^{c_{2}}/J_{z}=1.003$ is much slower than the rate 
at $J_{y}^{c}/J_{z}=0.4$ and $0.6$.

{\it Dimerized phase.}--We now concentrate on the dimerized phase in the vicinity of the isotropic point $J_x=J_y=J_z$, 
for which local order exists. 
With fixed $J_{y}/J_{z}=0.988$, SSB occurs when control parameter $J_{x}/J_{z}$ crosses the points 
$J_{x}^{c_{1}}/J_{z}=0.9799$ and $J_{x}^{c_{2}}/J_{z}=1.003$. 
To characterize the local order in the region between these two points, 
we consider dimerized local order parameter $\langle S_{i}S_{i+1}-S_{i+1}S_{i+2}\rangle$ \cite{Murray}.
Unfortunately, this definition only fits the system with SU(2) symmetry.
However, a systematic method based on tensor network representations to
derive local order parameters has been established \cite{Zhou2}.
Following this method, we analyse the combined dimerized local order parameters ${\boldsymbol D} =(D_{x},D_{y},D_{z})$, with
$D_{\alpha}=\langle S_{i}^{\alpha}S_{i+1}^{\alpha}-S_{i+1}^{\alpha}S_{i+2}^{\alpha}\rangle$. 
Fig.~\ref{fig4Nonlocal0988} shows plots of these dimerized local order parameters as
a function of control parameter $J_{x}/J_{z}$ with
truncation dimension $\chi=100$ for fixed $J_{y}/J_{z}=0.988$.
The dimerized local order ${\boldsymbol D}$ is clearly evident 
between $J_{x}^{c_{1}}/J_{z}=0.9799$ and $J_{x}^{c_{2}}/J_{z}=1.003$.
Fig.~\ref{fig4Nonlocal0988} clearly demonstrates the complementarity between the local and non-local order parameters.

{\it von Neumann entropy and central charge.}--To examine the nature of the QPT between SPt and dimerized phases we 
first discuss the definition of von Neumann entropy, as a measure of bipartition entanglement.
Consider the state $|\psi\rangle$ as being composed of two semi-infinite chains
$L(-\infty, \ldots, i )$ and $R(i+1, \ldots,+\infty)$, 
connected by the Schmidt decomposition coefficient $\lambda_{\alpha}$.
This implies $|\psi\rangle$ can be expressed as
$|\psi \rangle=\sum_{\alpha=1}^{\chi}\lambda_{\alpha}|\phi_{\alpha}^{L} \rangle|\phi_{\alpha}^{R} \rangle$,
where $|\phi_{\alpha}^{L} \rangle$ and $|\phi_{\alpha}^{R} \rangle$ are the Schmidt
bases of the two semi-infinite chains $L$ and $R$.
Consequently, the von Neumann entropy can be defined as \cite{Bennett}
$S=-Tr \, \rhoup_{L} \log \rhoup_{L}=-Tr \, \rhoup_{R}\log\rhoup_{R}$,
where $\rhoup_{L}=Tr_{R} \, \rhoup$ and $\rhoup_{R}=Tr_{L} \, \rhoup$
are the reduced matrices of the subsystems of $L$ and $R$, respectively,
with density matrix $\rhoup=|\phi\rangle\langle\phi|$.
For the semi-infinite chains $L$ and $R$ of iMPS, the von Neumann entropy
$S$ is written as
\begin{eqnarray}\label {entropydefinition}
S=-\sum_{\alpha=1}^{\chi}\lambda_{\alpha}^{2}\log\lambda_{\alpha}^{2}.
\end{eqnarray}

At a critical point in a 1D system, the semi-logarithmic scaling
of the von Neumann entropy versus truncation dimension $\chi$ follows from
conformal invariance, with scaling ruled by the central charge of the underlying conformal field theory.
In addition, the correlation length $\xiup$ of the iMPS exhibits a power scaling with truncation dimension $\chi$.
These two scaling relations can be written as \cite{KorepinCardy,Tagliacozzo,Pollmann3,Dai}
\begin{eqnarray}\label {scaling}
S_{\chi}\propto \frac{c\kappa}{6}\log{\chi}, \quad \xi_{\chi}\propto \xi_{0} \, \chi^{\kappa }.
\end{eqnarray}
Here $c$ denotes the central charge and $\kappa$ is a finite entanglement scaling exponent. $\xi_{0}$ is a constant.
For a given $\chi$, the correlation length $\xi$ can be obtained by the largest
and the second largest eigenvalues $D_{0}(\chi)$ and $D_{1}(\chi)$ of the transfer matrix, 
with $\xi_{\chi}=1/\log|D_{0}(\chi)/D_{1}(\chi)|$.
By making use of the relations (\ref{scaling}) one can obtain numerical estimates for the central charge 
 on the phase boundary between SPt phases
and between SPt and dimerized phases.

\begin{figure}[t]
\includegraphics[width=0.3\textwidth]{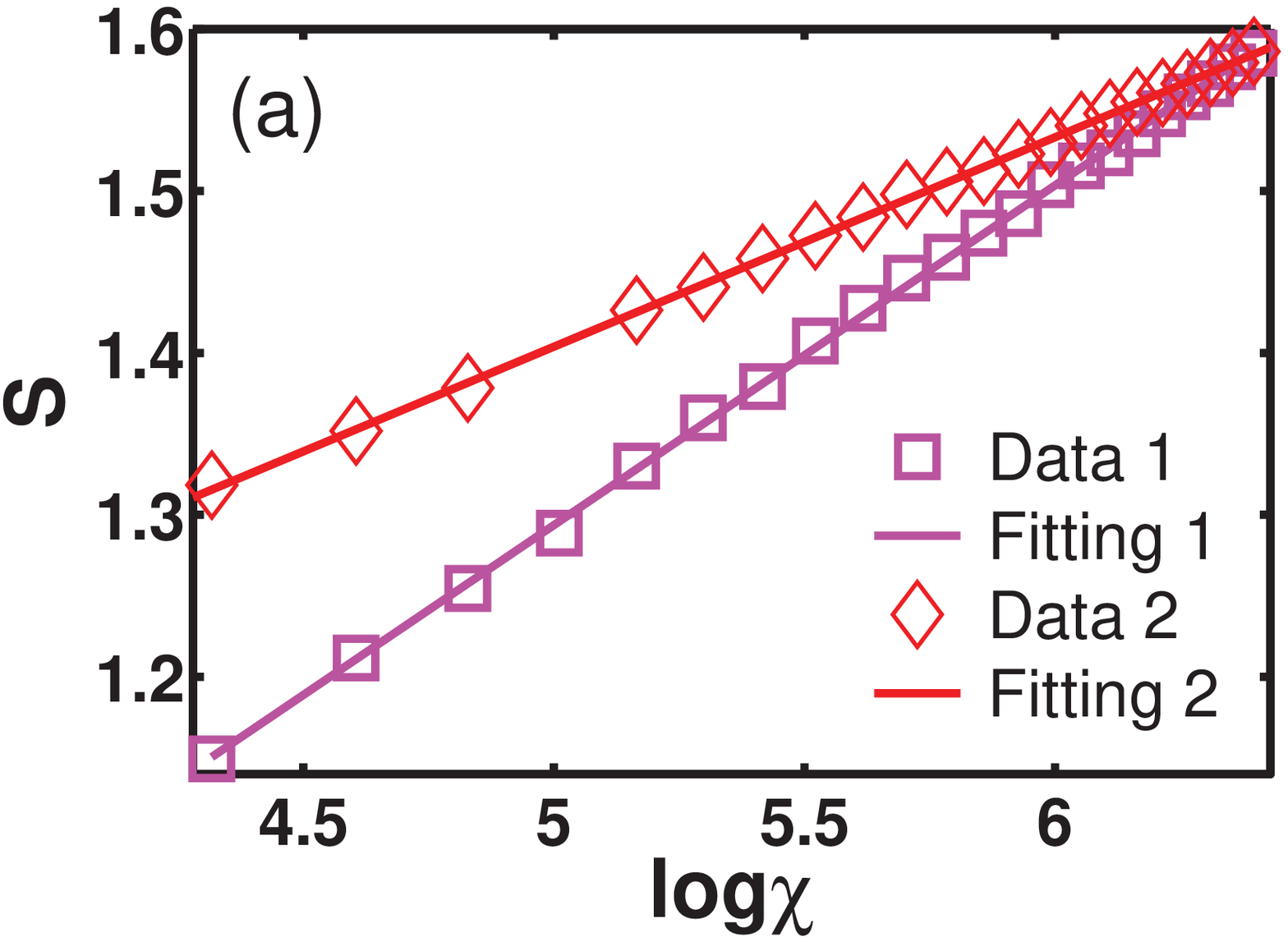}
\includegraphics[width=0.3\textwidth]{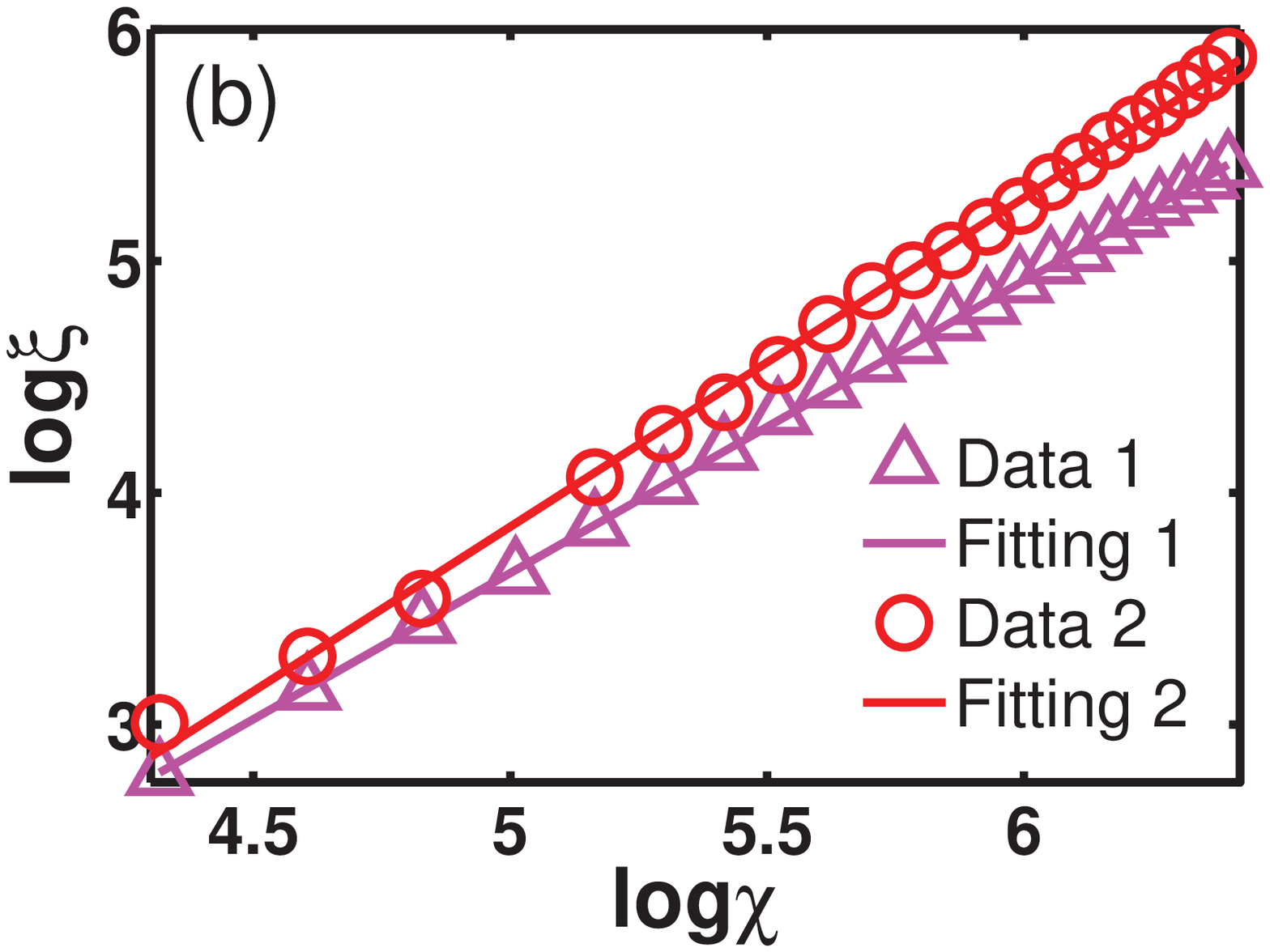}
\caption{The scaling of  (a) von Neumann entropy $S$ and (b) correlation length $\xi$ with the truncation dimension $\chi$.  
(i) For  fixed $J_{x}/J_{z}=0.4$ with $J_{y}^{c}/J_{z}=0.4$, shown in purple  and labeled as Fitting 1,
the central charge is estimated to be $c_{1}=0.997$.
(ii) For fixed $J_{y}/J_{z}=1$ with varying $J_{x}^{c}/J_{z}$,
shown in red and labeled as Fitting 2, the central charge is estimated to be $c_{2}=0.5463.$}
\label{fig5Sdiagonal}
\end{figure}

For this purpose, we choose $J_{y}^{c}/J_{z}=0.4$ with fixed $J_{x}/J_{z}=0.4$ and
choose $J_{y}/J_{z}=1$ with varying $J_{x}^{c}/J_{z}$. 
Fig.~\ref{fig5Sdiagonal} shows a corresponding plot of the von Neumann entropy and correlation length as a function
of truncation dimension $\chi$ ranging from 75 to 600. 
Both the von Neumann entropy and correlation length diverge with increasing truncation dimension $\chi$.
To extract the central charge, we use the fitting functions
$S_{\chi}=\frac{c\kappa}{6}\log{\chi}+a$ and $\log\xi_{\chi}=\kappa \log\chi+b$ and consider two cases.
(i) For the case of fixed $J_{x}/J_{z}=0.4$ with $J_{y}^{c}/J_{z}=0.4$, shown in purple in Fig.~\ref{fig5Sdiagonal} and labeled as Fitting 1,
the fitting coefficients are given by 
${c_{1}\kappa_{1}}/{6}=0.20968$, $a_{1}=0.24527$,  $b_{1}=-2.6563$, and
$\kappa_{1}=1.2619$.
The central charge is estimated to be $c_{1}=0.997$. This is consistent with a general argument that 
a phase transition between SPt phases belongs to the Gaussian universality class~\cite{Pollmann2}.
(ii) For the case of fixed $J_{y}/J_{z}=1$ with varying $J_{x}^{c}/J_{z}$,
shown in red in Fig.~\ref{fig5Sdiagonal} and labeled as Fitting 2,
the fitting coefficients are given by 
${c_{2}\kappa_{2}}/{6}=0.12929$, $a_{2}=0.75742$, $b_{2}=-3.2471$, and
$\kappa_{2}=1.4199$.
As a result, the central charge is estimated to be $c_{2}=0.5463$.  
This indicates that the phase transition falls into the Ising universality class, 
as anticipated from the fact that a $Z_2$ symmetry is spontaneously broken in the dimerized phase.

{\it Summary.}--We have investigated the nature of quantum SPt phases and quantum phase transitions in 
the spin-1 antiferromagnetic anisotropic biquadratic model  (\ref{ham}) by making use of quantum duality and symmetry transformations, 
along with iTEBD and iDMRG algorithms.  The concept of SPt phases, originally defined through the combined
operation of the site-centered inversion with the $\pi$-rotation around the
$y$-axis in  the spin space~\cite{Pollmann2}, is extended, in order to keep consistency with the duality
 transformations, which themselves are induced from the symmetric
group $S_3$ with respect to $x, y$ and $z$. 
Our results suggest the importance and potential generality of SPt phases in a classification of quantum states of matter.

The ground state phase diagram in Fig.~\ref{fig1duality}(b) has been determined by studying the principal regime, 
which can be mapped to the other five regions of the phase diagram via the quantum duality and symmetry transformations 
summarized in Table I. 
The phase boundaries are determined by calculating the non-local and local order parameters of the principal regime.
To illustrate our strategy, three sample lines, $J_{x}/J_{z}=0.4$, $J_{x}/J_{z}=0.6$ and $J_{y}/J_{z}=0.988$, are studied in detail.
The phase diagram is shown to be composed of four phases characterized by
the non-local order parameters ${\boldsymbol K}=(1,-1,-1)$, ${\boldsymbol K}=(-1,1,-1)$ 
and ${\boldsymbol K}=(-1,-1,1)$,  and a combined dimerized local order parameter. 
In addition, the von Neumann entropy and correlation length have been used to estimate the central charge $c=0.5463$
on the boundary between SPt and dimerized phases. 
This value is suggestive of the Ising-type universality class. 
The central charge value $c=0.997$ is extracted on the phase boundary between SPt phases, 
corresponding to the Gaussian-type universality class.
We have also identified three characteristic lines of factorized ground states, 
which are located in the SPt phases instead of a symmetry breaking phase, in sharp contrast to other known cases~\cite{factor1,factor2}.

{\it Acknowlegements.}- The work of M.T.B. has been supported by the National Natural Science Foundation of China Grant No. 11575037 and 
Australian Research Council Discovery Project DP180101040.


\end{document}